\documentstyle{mn}
\newcommand{\etal}{{\rm et al.~}}           % Defines et al.
\newcommand{\NH}{\mbox{${\rm N_H}$ }}       % Defines NH
\newcommand{\NHunits}{\mbox{$~{\rm cm}^{-2}$} }

\newcommand{\funit}{\mbox{erg~cm$^{-2}$~s$^{-1}$}}

\begin{document}
%\onecolumn
%\baselinestretch{2}
\title {Active Nucleus in a Poststarburst Galaxy : KUG~1259+280}
\author[Dewangan et al.]
{G. C. Dewangan$^1$,  K. P. Singh$^1$, Y. D. Mayya$^2$ and G.C. Anupama$^3$ \\
$~^1$Department of Astronomy \& Astrophysics, Tata Institute of Fundamental Research, Mumbai, India 400~005 \\
$~^2$Instituto Nacional de Astrofisica Optica y Electronica, Apdo Postal 51 y 216, 72000 Puebla, Pue., M\'EXICO \\
$~^3$Indian Institute of Astrophysics, Bangalore, 560~034 India}
\maketitle
%\slugcomment{Submitted to MNRAS, 2000 Feb 22}
\begin{abstract} 
We report the discovery of an active nucleus in a poststarburst 
galaxy, KUG 1259+280, based on its X-ray and optical characteristics.
The X-ray source in KUG 1259+280 was detected during 
a search for ultra-soft sources in the ROSAT pointings.
High resolution X-ray imaging observations with $ROSAT$ HRI show that 
X-ray emission from KUG~1259+280 is unresolved. 
X-ray emission from 
KUG~1259+280 is highly variable, an episode in which X-ray intensity 
changed by a factor $\sim 2.5$ within $\sim$ 1300~s has been detected. 
$ROSAT$ PSPC spectra of this galaxy is found to be well represented by a 
steep power-law of photon index, $\Gamma_{X}\sim4.25$, and 
a change in the absorbing column within $\sim$ 1~d is indicated. 
The rest frame intrinsic X-ray luminosity of KUG~1259+280 is found to 
be $\sim3.6-4.7\times10^{42}{\rm~erg~s^{-1}}$ similar to that of low 
luminosity Seyfert galaxies. Mass of the central 
massive object within KUG~1259+280 is estimated to be in the range 
$10^{5}-10^{7}M_{\sun}$. Optical spectrum of the nuclear region of the 
galaxy is complex and shows Balmer absorption and collisionally excited 
lines of [O III], and [N II].  The presence of 
forbidden emission lines and the absence of Balmer emission 
lines in spectrum of KUG~1259+280 could be due to photoionization by a diluted 
power-law continuum modified by enhanced stellar absorption due to a 
poststarburst event. Estimated Balmer line strengths free of stellar 
absorptions and forbidden line strengths indicate the nucleus of 
KUG~1259+280 to be LINER-like in nature. However, the low-ionization forbidden line 
[O I]$\lambda$6300 usually present in LINER spectra, is not detected in the spectrum 
of KUG~1259+280. X-ray characteristics -- variability, point-like appearance, luminosity and 
steepness of spectrum indicate that nucleus in KUG~1259+280 is active and perhaps
like that of narrow-line Seyfert type 1 galaxies. 
\end{abstract}

\section{Introduction}
Recent optical (Ho, Filippenko, \& Sargent 1997a,b) and X-ray surveys 
(Grupe \etal 1998) are finding  that nuclei of many nearby galaxies are 
 active.  Such activities manifest themselves
through continuum and emission-line properties.  Characterizing  these
properties helps us to measure the luminosity functions and evolution of 
different sub-types, and understanding the relationship between different
types  of activities, for example,  between nuclear starbursts and 
active galactic nuclei (AGNs), and thus assist in creating unified models for
active galaxies.  Weedman (1983) has argued that 
starburst events in the nuclei of galaxies would evolve rapidly
into compact configurations as dynamically distinct entities. Perhaps the 
compact configuration further collapses to form a single massive black 
hole generally favored as the central engine for AGNs (Miller 1985).
On the other hand, Goncalves, Veron-Cetty \& Veron (1999) suggest that
perhaps it is nuclear activity that triggers circumnuclear star-formation.
However, evidence for either case is fragmented at present.
In this connection, it is very important to study AGNs in starburst 
and poststarburst galaxies. In this paper, we report new X-ray and optical
observations of a poststarburst galaxy KUG~1259+280.

The galaxy, known as KUG~1259+280 (Takase \& Miyauchi-Isobe 1985), is a 
poststarburst galaxy that appears to be in a small group of 4 galaxies 
within an arcmin, and is a member of the Coma cluster, where it is known 
variously as D-61 (Dressler 1980) or GMP~1681 (Godwin, Metcalfe, \& Peach 
1983).  KUG~1259+280 is the brightest among these 
4 galaxies (Takase \& Miyauchi-Isobe 1985) with $B=15.88{\rm~mag}$
and has a heliocentric radial velocity of $7102\pm50{\rm~ km s^{-1}}$ (Caldwell \& Rose 1997).
From ground-based imaging, KUG~1259+280 has been classified as an SO galaxy. 
However, the high resolution images of KUG~1259+280 taken with the 
{\it Hubble Space Telescope} have revealed this galaxy to be an edge-on 
disk galaxy with a very bright unresolved nucleus (Caldwell, Rose, \& 
Dendy 1999, hereafter CRD). Two radial dust lanes in KUG~1259+280, to the 
east of the centre and about $1\arcsec$ away, have also been reported by CRD. 
An unusual bulge that is boxy or peanut-shaped and, additionally, has an 
X morphology has been found by CRD.  They also report that the central 
($0.16\arcsec$ or 74~pc for $H_{0}=75{\rm~km~s^{-1}~Mpc^{-1}}$ ) 
luminosity of KUG~1259+280 is $M_{B} = -16.3$ (CRD) which is higher than 
any of the star clusters found in the merging galaxies studied by 
Whitmore \etal (1997) or by Carlson \etal (1998) and is also higher than 
the luminosity of nuclear star clusters found in normal early-type galaxies 
studied by Lauer \etal (1995). The nucleus is about 0.2 mag bluer than that 
of the surrounding bulge (CRD). In their spectroscopic study of galaxies in 
the Coma cluster, CRD have found that KUG~1259+280 is a very strong 
poststarburst galaxy with poststarburst age of $\sim0.5{\rm~ Gyr}$ which is shorter than the
typical value of 1 Gyr for similar galaxies in Coma, thus indicating that the
starburst in this galaxy had happened more recently as compared to other
such galaxies in Coma.  Based on the ratio of flux of 
[N II]$\lambda\lambda6548,6583$ and H$\alpha$, CRD have concluded that 
the emission line spectrum of KUG~1259+280 to be arising due to LINER-like 
nuclear activity.

\begin{table*}
\caption{Basic parameters of KUG~1259+280}
\begin{flushleft}
\begin{tabular}{l}
\hline
Other names$^1$ :  CGCG 1259.6+2803, CG 0959, FOCA 0357, PGC 044947, \\
~~~~~~~~~~~~~~~~~~~~~ RX J1301.9+2746, RX J1301.9+2747, Coma-D~61, GMP~1681. \\
Morphological type : SO  \\
Position$^2$ : $\alpha(J2000) = 13^{h}~02^{m}~00.1^{s}$; $\delta(J2000) = +27^{\circ}~46^{\arcmin}~57^{\arcsec}$  \\
Redshift$^3$ : $z$ = $0.02369\pm0.00017$  \\
Magnitude$^3$ : $B$ = $15.88$ \\
Color$^3$ : $B-R$ = 1.63 \\
 \hline
$~^1$Nasa Extragalactic Database (NED) \\
$~^2$Doi \etal (1995) \\
$~^3$Caldwell \& Rose (1997)\\
\end{tabular}
\end{flushleft}
\end{table*}

The galaxy, KUG~1259+280, was first detected as a 
serendipitous $EXOSAT$ X-ray source, EXO~125938+2803.1, in the 
region of Coma cluster by Branduardi-Raymont \etal (1985).
The galaxy was also identified as an X-ray source in the $EXOSAT$ 
high Galactic latitude survey by Giommi \etal~(1991). 
X-ray emission from KUG~1259+280 was again detected  during
the $ROSAT$ pointings and was reported by Singh \etal (1995) to be an ultra-soft X-ray source -- WGA J1301.9+2746 (RX J1301.9+2746) (White, Giommi, \& Angelini 1994).
Based on this association,  
Singh et al. (1985) obtained soft X-ray luminosity ($L_{X}$) of 
$\sim$3 $\times$ 10$^{42}$ erg s$^{-1}$ for KUG~1259+280 which is similar to 
that of a low-luminosity AGN. The basic parameters of KUG~1259+280 are 
summarized in Table 1.

In this paper, we present a detailed analysis of X-ray data
on the  ultra-soft X-ray source -- WGA J1301.9+2746 (RX J1301.9+2746), taken
from the $ROSAT$ archives.  
The data taken on 5 different days of observations during 1991 to 1995 have
not been presented before.  We confirm the identification of 
the X-ray source with KUG~1259+280 based on high resolution X-ray imaging 
data. We present X-ray light curves and
spectra  of KUG~1259+280 taken on 3 occasions. We  also present our new optical 
spectroscopic observations of the galaxy, and discuss its peculiar nature.

Throughout the paper, luminosities are calculated assuming a Hubble 
constant of $H_{0}=75{\rm~km~s^{-1}~Mpc^{-1}}$ and a deceleration 
parameter of $q_{0}=0$.

\section {Observations}
\subsection{$ROSAT$ X-ray Observations}
The region of the sky containing this source was observed three times with 
the $ROSAT$ (Truemper et al. 1983) Position Sensitive Proportional
Counter (PSPC) and twice with the High Resolution Imager (HRI)
(Pfeffermann et al. 1987). 
The PSPC observations were carried out during 1991 June 16-19.
Coma cluster was the target source in the PSPC observations, whereas
KUG~1259+280 itself was targeted in the HRI observations of 1995 June 5.   
The galaxy was again in the field of view of the HRI observations
during 1994 June 25, when NGC~4921 and NGC~4923 were targeted. The details of $ROSAT$
observations are given in Table 2. The off-set of KUG~1259+280 from the
field centre is also listed in Table 2.
ROSAT X-ray data corresponding to the above observations were obtained 
from the public archives maintained at the High Energy Astrophysics Science
Archive Research Center (HEASARC) in USA, and analysed by us.  An X-ray 
source centered on the position of KUG~1259+280 was detected in all 
the observations. It is offset from the centre of the field of view of the 
PSPC by $\sim0.5^\circ$, however, it was
was not obstructed by the PSPC window support structure in the observations 
identified by the sequence numbers RP800006N00 (1991 June 16) and 
RP800013N00 (1991 June 18). In the 1991 June 17 observation (RP80005N00),
the source is somewhat closer to the PSPC window structure, however, it does 
not appear to be obstructed in any significant way by the window structure
even though the point spread function of the PSPC and the telescope system is
very broad at this position.
\begin{table*}
\caption{Details of ROSAT observations of RX~J1236.9+2656}
\begin{flushleft}
\begin{tabular}{lllllll}
 \hline
Serial & Sequence & Instrument & Off-set & Start Time & End Time & Exposure  \\
No. & No. &  & arcmin & Y, M, D, UT & Y, M, D, UT & Time (s)  \\
\hline
1. & RH900633N00 & HRI  & 0.16  &1995 06 05 22:36:48 &1995 06 06 19:53:43 & 5616   \\
2. & RH701579N00 & HRI  & 9.8   &1994 06 25 12:24:51 &1994 07 07 04:29:39 & 17765   \\
3. & RP800006N00 & PSPC & 29.8  &1991 06 16 22:46:34 &1991 06 17 22:44:23 & 21545 \\
4. & RP800005N00 & PSPC & 31.8  &1991 06 17 22:44:52 &1991 06 18 22:43:06 & 21140  \\
5. & RP800013N00 & PSPC & 28.6  &1991 06 18 22:43:30 &1991 06 19 22:41:59 & 21428  \\
\hline
\end{tabular}
\end{flushleft}
\end{table*}

\subsection{Optical Observations}
Optical V band image of KUG~1259+280 was obtained at the Vainu Bappu 
Telescope (VBT), Kavalur, India on the night of 1996 March 13. The observation 
was carried out using a $1024\times1024$ CCD chip placed at the prime focus of the 2.3~m 
reflector. The pixel size of $24{\rm~\mu m~pixel^{-1}}$ of the CCD gives a scale of $0.61\arcsec$ and a total field of view $10.4\arcmin \times10.4\arcmin$. 

Spectroscopic observations of KUG~1259+280 were carried out under photometric conditions on the night of 2000 Feb 02 at
the Cassegrain focus of 2.12-m telescope of the Guillermo Haro Observatory (GHO), Cananea, Mexico.
The Boller and Chivens Spectrograph with TEX $1024\times1024$ CCD and $150{\rm ~l~mm^{-1}}$ grating was used. The slit position angle was $48^\circ$ which passes roughly along the major axis. The slit width was $2\arcsec$. The airmass of the observations was close to 1.0.
The instrumental resolution (FWHM) was $\sim10{\rm~\AA}$. Three spectra of the central
region of KUG~1259+280 were acquired in order to increase the
signal-to-noise ratio and to reduce unwanted cosmic ray hits.
The exposure times were $1800$~s for each of the three exposures.
To correct for the bias level and nonuniform
response of the pixels in the CCD, bias frames and dome flat
frames were also acquired on the same night.
For the purpose of wavelength calibration, comparison (He-AR) spectra
were obtained. The spectrophotometric
standard star HR~4963 was observed (2 exposures) on the same night
to flux calibrate the object spectra.

Spectroscopic observations ($2$ exposures of $2400{\rm~s}$ each) of KUG~1259+280 were also carried out on the night of 1999 Feb 26 at the Cassegrain focus of the VBT using OMR
spectrograph (Prabhu, Anupama, \& Surendirath 1998) with a slit of width $2.8\arcsec$, $600{\rm ~l~mm^{-1}}$ grating and $1024\times 1024$ CCD camera. The instrumental resolution (FWHM) was $\sim7{\rm~\AA}$, and the slit
position angle was $0^\circ$. The sky conditions were close to photometric. However, the spectra obtained were of poor signal-to-noise ratio due to poor reflectivity of the primary. Because of the wider wavelength range covered and better signal-to-noise ratio
obtained at GHO, we present only the results obtained from the GHO.

\section{Analysis \& Results}
\subsection{Image Analysis}
The analysis of X-ray images has been carried out using the 
PROS\footnote[6]{The PROS software package provided by the $ROSAT$  Science 
Data Center at Smithsonian Astrophysical Observatory.} software package. 
X-ray images were extracted from all the observations listed in Table 2. The high 
resolution (central full width half maximum of $\sim4\arcsec$) images, 
obtained with HRI, were smoothed by convolving with a Gaussian of 
$\sigma$=$5\arcsec$. 
The PSPC images were smoothed with a Gaussian of $\sigma=10\arcsec$. 
X-ray contours of smoothed HRI and PSPC images have been overlaid on 
the optical $V$ band image and  are shown 
in Figures 1 and 2, respectively. It is clear from Fig. 1 that the 
X-rays are centered on the galaxy, KUG~1259+280. 
No other X-ray source, within the angular spread comparable to the point spread 
function of $ROSAT$ PSPC, was seen in either of the two HRI observations. 
Therefore,  X-ray emission from KUG~1259+280 is not contaminated by 
emission from any other source.  

\begin{table*}
\centering
\caption{Position and count rates of KUG~1259+280 derived from $ROSAT$ observations.}
\begin{flushleft}
\begin{tabular}{lcccc}
\hline
Date of & Instrument & \multicolumn{2}{c}{Position} & Source Count Rate \\
observation &        & $\alpha(2000)$ & $\delta(2000)$ & $10^{-2}{\rm~count~s^{-1}}$  \\
\hline
1995 June 06 & HRI   & $13^{h}~01^{m}~59.9^{s}$ & $+27^{\circ}~47{\arcmin}~00.8{\arcsec}$ & $2.4\pm0.24$ \\

1994 June 25 & HRI   & $13^{h}~02^{m}~00.2^{s}$ & $+27^{\circ}46^{\arcmin}~59.7{\arcsec}$ & $2.3\pm0.13$ \\

1991 June 16 & PSPC & $13^{h}~02^{m}~00.0^{s}$ & $+27^{\circ}~47\arcmin~03.9\arcsec$ & $17.7\pm0.43$ \\

1991 June 17 & PSPC & $13^{h}~02^{m}~00.8^{s}$ & $+27^{\circ}~47\arcmin~04.0\arcsec$ & $17.3\pm0.44$ \\

1991 June 18 & PSPC  & $13^{h}~02^{m}~00.5^{s}$ & $+27^{\circ}~46\arcmin~59.7\arcsec$ & $16.6\pm0.40$ \\

1990 July 11 (RASS$^1$)& PSPC  & $13^{h}~01^{m}~58.9^{s}$ & $+27^{\circ}~47\arcmin~08.5\arcsec$ & $12.43\pm1.95$ \\
\hline
$^1${\em IRAS} All Sky Survey (Voges \etal 1999). \\
\end{tabular}
\end{flushleft}
\end{table*} 

We have estimated count rate of KUG~1259+280 from the PSPC observations. 
Due to the large off-axis angle ($\sim30\arcmin$) of KUG~1259+280 from 
the field centre in each of the PSPC observations, it is necessary to 
correct the PSPC images for exposure and vignetting. For this reason, we 
have created exposure map for each of the PSPC observation using 
appropriate devignetted detector map obtained from HEASARC. Exposure 
corrections to the PSPC images were carried out using the XIMAGE software.
Source count rates were estimated from square boxes centered at the source 
position and after subtracting the background estimated from square annuli 
centered at the source position but outside the source region. The source 
box size was optimally selected to maximize the signal-to-noise ratio. 
The count rates thus estimated are corrected for exposure and vignetting 
and are given in Table 3. The X-ray intensity of KUG~1259+280 remained 
almost steady during the 3 pointed PSPC observations. X-rays from KUG~1259+280 
were also detected in 1990 during the $ROSAT$ All Sky Survey (RASS) 
(Voges \etal 1999). 
X-ray intensity of KUG~1259+280 obtained from the $ROSAT$ All-Sky Survey Bright 
Source Catalogue (Voges \etal 1999) is given in Table 3, and is about $25\%$ less
than that during the pointed mode observations in 1991.

%The source counts from the unsmoothed PSPC images were obtained from  
%circles of radii $3.25\arcmin$
%centered on the peak position, and after subtracting the background 
%estimated from three nearby circular regions away from the Coma cluster 
%with their centres about $6.5\arcmin$ away from the source. 
%This was done using the $imcnts$ program in 
%the PROS software package.  The large radius for source extraction
%was necessitated due to the broadening  of the point spread function (psf)
%at  large off-axis angle from the centre of the field of view of the
%telescope.  The optimal extraction radius for the source was selected by 
%changing the radius and finding the maximum in the total source counts with best signal-to-noise ratio.
%The background estimates were checked by changing the number of circles 
%used for background determinations and no significant change was found.
%The count rates thus estimated are shown in Table 2.
%The X-ray intensity in the two observations carried out on 1991 June 16 
%and 1991 June 18 was, thus, almost unchanged. The source count rate in 
%the observation of 1991 June 17 is, however, slightly lower than in the other two 
%PSPC observations but is not significant and could perhaps be due to a
%slight obstruction of the extended wings of the point source by the PSPC 
%window structure (see also \S2.3).

The source counts from the HRI images, extracted from the data observed on 1994 
June 25 and 1995 June 5, were estimated from circles of radii $30\arcsec$, 
and $25\arcsec$, respectively, after subtracting the background counts estimated from 
annular regions centered at the peak positions with inner and outer radii of
$60\arcsec$ and $120\arcsec$. The optimal extraction radii were chosen from 
radial profiles for the KUG~1259+280 to include the maximum signal and minimum 
background noise. The background estimates were checked by changing the width 
of the annuli and no significant change in total counts was noticed. 
The HRI count rates are given in Table 3. The count rates in the two HRI observations are unchanged. 

{\begin{figure*}
\vskip 12cm
\includegraphics{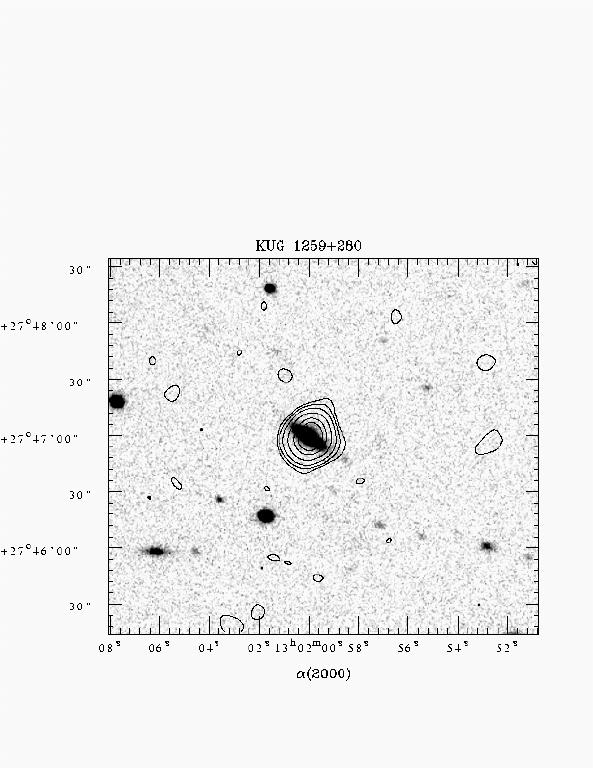}
\caption {Contours of $ROSAT$ HRI image of KUG~1259+280 overlaid on the $V$ band
optical image observed with 2.3-m Vainu Bappu Telescope, Kavalur, India.
The contours are shown at $4\%$, $6\%$, $10\%$, $20\%$, $40\%$, $60\%$, and $80\%$ of
the peak intensity ($0.3153{\rm ~counts~pixel^{-1}}$). Background level is $0.0056\pm0.0004{\rm ~counts~pixel^{-1}}$. The X-ray contours have been generated from the HRI
image after smoothing by a Gaussian of $\sigma=5\arcsec$. The HRI image was created from the data observed on 1994 June 25 (exposure time = 17765~s).}

\end{figure*}}

{\begin{figure*}
\vskip 14cm
\includegraphics{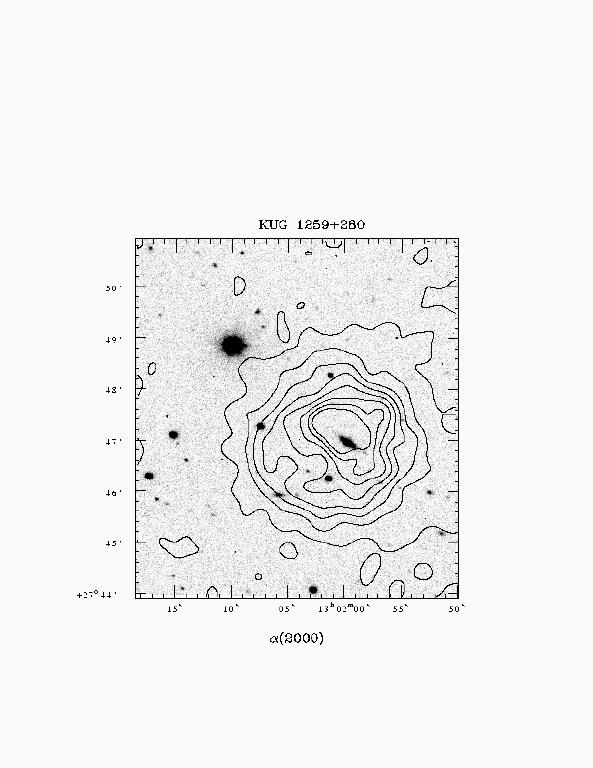}
\caption{Contours of $ROSAT$ PSPC image of KUG~1259+280 overlaid on the $V$ band
optical image observed with 2.3-m Vainu Bappu Telescope, Kavalur, India.
The X-ray contours are shown at levels $10\%$, $20\%$, $30\%$, $40\%$, $50\%$, $60\%$, $70\%$, and $80\%$ of the peak intensity ($0.0471{\rm~count~pixel^{-1}}$).
The background level is $0.0035\pm0.0001{\rm~count~pixel^{-1}}$. X-ray contours
were derived from X-ray image after smoothing by a Gaussian of $\sigma=10\arcsec$.
The X-ray image was  generated from the data observed on 1991 June 18.}

\end{figure*}}

In order to find the spatial extent of KUG~1259+280 in X-rays, we have generated
a radial intensity profile of the X-ray source in KUG~1259+280 from raw $ROSAT$ 
HRI images extracted from the 1995 June 5 data, and
compared it with the point response function of the telescope and detector.
We used this HRI observation because the source is centered on the axis, even though
the exposure time is smaller than in the other HRI observation. 
This was done because the HRI point response function broadens at off-axis positions.
The radial profile of KUG~1259+280 was generated by averaging azimuthally
in radial bins of $2\arcsec$ from the raw HRI images. Background intensity
was estimated from an annular region of width $60\arcsec$ centered at the
galaxy, $2.5\arcmin$ away from the galaxy centre. The radial profile 
thus generated was normalized by the intensity at a  radius of $1\arcsec$ and
is shown in Figure 3. 
The point response function of the
$ROSAT$ telescope and HRI was generated using the analytic form of the  on-axis
point response function (David \etal 1993). 
As can be seen in Fig. 3,
KUG~1259+280 has not been spatially resolved by $ROSAT$ HRI. It should be
noted that the slight excess over the theoretical PRF around radii of
 $5\arcsec$ to $10\arcsec$ does not imply that the galaxy is spatially extended.
The work of Lehmann \etal (1998) has proved that the more realistic
re-calibrated HRI PRF differs slightly but significantly from the theoretical HRI PRF
in the range $10\arcsec-30\arcsec$. Thus we conclude that there is no evidence for 
extended X-ray emission from KUG~1259+280 based on the $ROSAT$ HRI observations.

{\begin{figure*}
\vskip 11cm
\includegraphics{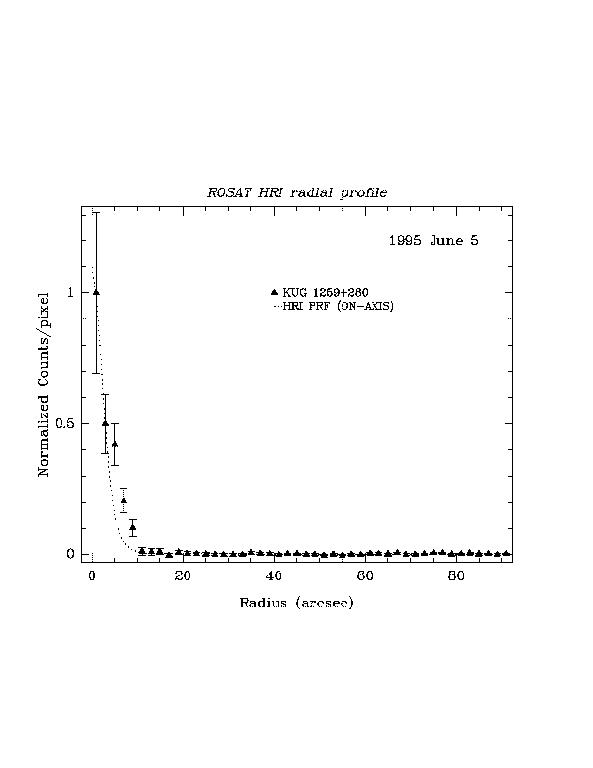}
\caption { Radial HRI intensity distribution of KUG~1259+280 compared with the theoretical PRF of the $ROSAT$ HRI. The radial profile of KUG~1259+280 has been derived from the data
obtained on 1995 June 5, with minimum off-set ($0\arcmin.16$) from the centre of the field.}
\end{figure*}}

\subsection{X-ray Timing}
In order to investigate the time variability of the soft X-ray emission from
KUG~1259+280, we have extracted light curves from the three $ROSAT$ PSPC
observations. The light curves were extracted
using the `xselect' package with time bins of 400~s and in the full energy
band of $ROSAT$ PSPC containing all the X-ray photons. 
For each PSPC observation, source$+$background light curve was extracted from a
circle of radii $3.25\arcmin$
centered on the peak position, while background light curve was extracted
from three nearby circular regions away from the Coma cluster
with their centres about $6.5\arcmin$ away from the source. The large radius for source extraction circle
was necessitated due to the broadening  of the point spread function (psf)
at  large off-axis angle from the centre of the field of view of the
telescope. 
The background 
subtractions were carried out after appropriately scaling the background 
light curves to have the same area as the source extraction area. 
The light curves  of KUG~1259+280 thus generated are shown 
Figures 4(a), 4(b), and 4(c) for the PSPC observations carried out on 1991 
June 16, 17, and 18 respectively. As can be seen in the Figures 4(a) and 4(b), X-ray
emission from KUG~1259+280 is steady during the observations on 1991 June 16 and 17.
However, the galaxy shows a remarkable X-ray variability on 1991 June 18 when an event with peak
X-ray intensity of 2.5 times the average value that subsequently
declined to the average value within about 1300~s, was recorded.
It should be noted that the variability amplitude and the time scale could be 
larger than seen, since the observation does not cover the full time span of 
the variable event due to Earth occultation. After the decline, X-ray intensity 
from KUG~1259+280 remains almost steady. To investigate any change in the 
hardness ratio during the flare or outburst, we generated two light curves 
of KUG~1259+280 in the energy bands 0.1--0.3 keV and 0.3--1.5 keV. From these 
light curves, we calculated hardness ratio as a function of time and found 
no changes in the hardness ratio during the outburst or flare. Since most
of the X-ray photons are concentrated below 0.5 keV, we  confirmed the above
result by calculating hardness ratios using light curves extracted in the 
energy bands 0.1--0.35 keV, 0.35--1.5 keV and 0.1--0.4, 0.4--1.5 keV. 

{\begin{figure*}
\vskip 18cm
\includegraphics{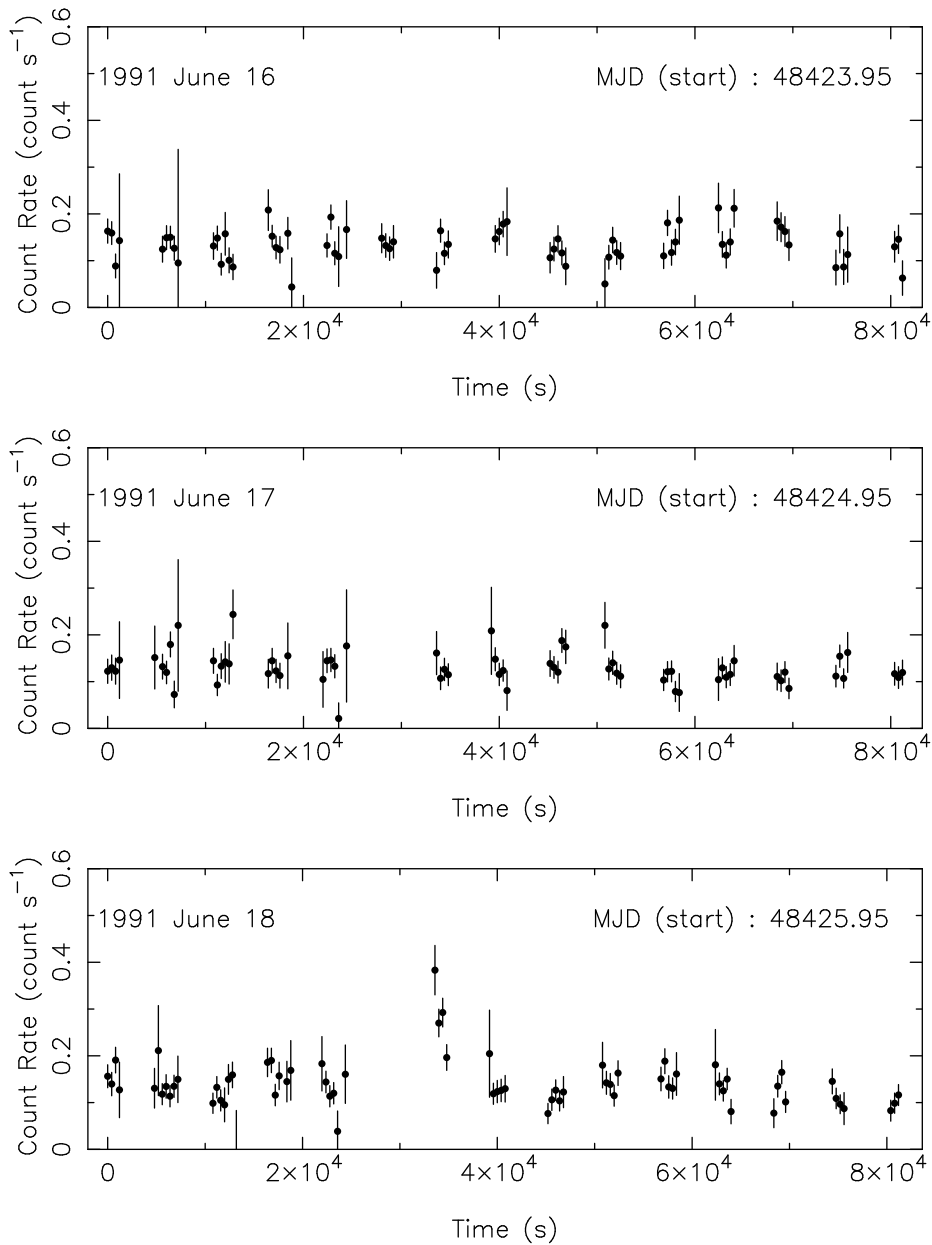}
\caption{ $ROSAT$ PSPC light curves of KUG~1259+280 observed on (a) 1991 June 16, (b) 1991 June 17, and (c) 1991 June 18. The time bin size is 319~s. Short term variability is seen on 1991 June 18.}
\end{figure*}}

In order to find out any possibility of the presence of long-term variation of soft X-ray emission from KUG~1259+280, we have searched the literature for observations of the galaxy with other X-ray observatories. The galaxy was not observed with the Einstein observatory (Branduardi-Raymont \etal 1985). The $EXOSAT$ LE count rate of KUG~1259+280 observed on 1985 June 17 is $7.5\times10^{-3}{\rm~count~s^{-1}}$ (Giommi \etal 1991). The $EXOSAT$ LE count rate of KUG~1259+280 corresponds to an observed flux of $\sim4\times10^{-13}{\rm~erg~cm^{-2}~s^{-1}}$ in the energy band of 0.05-2.0~keV using the spectral parameters measured with the $ROSAT$ PSPC ($\NH=1.4\times10^{20}{\rm~cm^{2}}$, $\Gamma_{X}=4.25$) (see $\S 3.3$). Using W3PIMMS, this flux is found to be equivalent $ROSAT$ PSPC count rate of $\sim0.12$, which is similar to that obtained from the RASS in 1990 but smaller by $\sim25\%$ than during the pointed observation in 1991. We have also converted
 the  $ROSAT$ HRI count rates of KUG~1259+280 to the PSPC count rates using the best-fit spectral parameters, for the purpose of comparison. The equivalent PSPC count rates are $(15.2\pm1.5)\times10^{-2}{\rm~count~s^{-1}}$ and $(15.1\pm0.8)\times10^{-2}{\rm~count~s^{-1}}$, for the HRI observations carried out on 1995 June 6 and 1994 June 25, respectively. Thus X-ray emission from KUG~1259+280 is almost steady during the pointed PSPC and HRI observations. 
%As can be seen from Table 3, X-ray intensity of KUG~1259+280 increased by about $\sim25\%$ within a year between RASS to PSPC pointed observations.
In summary, X-ray intensity of KUG~1259+280 was same in 1990 observations as in the 1985 observations, and then increased by $\sim25\%$ in 1991.

\subsection{X-ray Spectra}
Photon energy spectra were accumulated from the PSPC events, separately for the 
three observations shown in Table 2,  and using the same source 
and background regions as stated above.
The $ROSAT$ PSPC pulse height data obtained in 256 pulse height channels
were re-grouped  by adding counts in 8 consecutive channels to improve 
the statistics.  The X-ray spectra from the three observations thus obtained  
are shown together in the upper panel of Figure 5. 
Bad channels have been ignored.  

We used the XSPEC (Version 10.0) spectral analysis package to
fit the data with spectral models.   This requires a knowledge
of the response of the telescope and the detector. 
Using the source spectral files and information about the source
off-axis angle contained within them, we generated an auxiliary response file of the
effective area of the telescope, by using the program $pcarf$ in the
FTOOLS (version 4.2) software package.  This program uses the
available off-axis calibration of the telescope, and the amount of
scattering of X-ray photons and their dependence on energy are taken
into account.  Appropriate response matrix, provided
by the ROSAT GOF at HEASARC, was used to define the energy response of 
the PSPC.  The $ROSAT$ PSPC spectra, shown in Fig. 5, were used for fitting
spectral models.  The spectra from the three observations were first fitted 
separately with a redshifted power-law model with photon index, $\Gamma_{X}$, 
and absorption due to an intervening medium with absorption cross-sections as 
given by Balucinska-Church and McCammon (1992) and using the method of
$\chi^{2}$-minimization. The results of this fitting and the best-fit 
spectral model parameters are shown in Table 4.
This simple model is a good fit to the data of 3 days as evidenced by 
the minimum reduced $\chi^{2}$ values ($\chi^{2}_{\nu}$).
The spectra corresponding to the first and the second 
observations are consistent with each other in the sense that both the 
best-fit photon indices and the column densities of the absorbing material are 
similar within the errors.  The absorber column densities are similar (within errors) to 
the 21-cm value  (\NH=9.5$\times10^{19}$\NHunits) measured 
from radio observations in this direction (Dickey \& Lockman 1990), indicating that all the X-ray 
absorption is due to matter in our own Galaxy.  Therefore, we have also fitted the
power-law models to these spectra after fixing the neutral hydrogen column 
density to the Galactic value. The best-fit spectral parameters, in this case, 
are also given in Table 4. The best-fit minimum $\chi^{2}_{\nu}$ and the power-law 
index do not change significantly from those obtained while varying the 
$N_{H}$, as can be seen in Table 4. The photon indices thus obtained are, however,
better constrained and are $3.80_{-0.24}^{+0.14}$ and $4.0_{-0.3}^{+0.2}$ for 
the spectra observed on 1991 June 16 and 1991 June 17, respectively. 
The errors quoted, here and below, were calculated at the 90\% confidence 
level based on $\chi^2_{\rm min }$+2.71.
Based on the best fit model parameters, the observed X-ray flux from the source 
is estimated to be $6.3\times$10$^{-13}$ erg cm$^{-2}$ s$^{-1}$ for the 
observation on 1991 June 16, and $5.6\times$10$^{-13}$ erg cm$^{-2}$ s$^{-1}$ 
for the observation on 1991 June 17. 
We have also estimated the intrinsic flux of KUG~1259+290 by setting \NH to zero. 
The intrinsic flux of KUG~1259+280 in the energy band 0.1--2.0 keV 
is $2.4\times$10$^{-12}{\rm~ erg~cm^{-2}~s^{-1}}$ on 1991 June 16, and 
$2.3\times10^{-12}{\rm~ erg~ cm^{-2}~s^{-1}}$ on 1991 June 17. 

\begin{table*}
%\centering
\caption{Best-fit Model spectral parameters for KUG~1259+280}
%\begin{flushleft}
\begin{tabular}{llllllll}
\hline
Date of & Model$^a$ &  \NH & Excess \NH & $\Gamma_{X}$ & Norm$^b$ & $f_{X}^{c}$ &  $\chi^{2}_{\nu}/\nu^{~d}$ \\
observation & & $10^{20}{\rm ~cm^{-2}}$  & $10^{20}{\rm ~cm^{-2}}$ &  or $kT$ (keV) & & &  \\
\hline
1991 June 16 & phabs(zpl) & $1.1_{-0.25}^{+0.25}$ & - & $4.0_{-0.4}^{+0.3}$ &$4.1_{-1.2}^{+2.0}$ & 6.25 &  0.997/29 \\
\\
            & phabs(zpl) & $0.95 $            & - & $3.8_{-0.24}^{+0.14}$ & $4.5_{-0.9}^{+2.0}$ & 6.29 & 0.999/30 \\
\\
1991 June 17 &phabs(zpl)& $1.15_{-0.21}^{+0.37}$ & - & $4.1_{-0.3}^{+0.75}$ & $3.3_{-1.9}^{+1.4}$ & $5.68$ & 0.960/29 \\
\\
            &phabs(zpl) & 0.95                 & - & $4.0_{-0.3}^{+0.2}$ & $3.2_{-1.8}^{+2.0}$ & $5.56$  & 1.011/30 \\
\\
1991 June 18 &phabs(zpl) & $1.9_{-0.25}^{+0.45}$ &-& $4.5_{-0.3}^{+0.6}$ & $3.5_{-1.6}^{+1.1}$ & $6.5$ & 0.985/22 \\
\\
            &phabs(zpl) & $0.95$                &-& $3.7 $ & 5.2 & - &  2.427/23 \\
\\
            &phabs(zphabs)zpl & 0.95 & $1.0_{-0.3}^{+0.5}$ & $4.57_{-0.32}^{+0.56}$ & $3.4_{-1.5}^{+1.2}$ & 6.47 & 0.979/22 \\
\\
1991 June 16,17,18  &phabs(zpl) & $1.4_{-0.2}^{+0.2}$ &-& $4.25_{-0.22}^{+0.25}$ &$3.6_{-0.9}^{+0.9}$ & 6.1 & 1.119/86 \\
\\
       &phabs(zphabs)zpl &0.95 & $0.5_{-0.2}^{+0.2}$ & $4.25_{-0.22}^{+0.25}$ & $3.6_{-0.9}^{+0.9}$ & 6.1 & 1.119/86 \\
\\
         &phabs(zbbody) &0.95 &- & $0.055_{-0.006}^{+0.005}$ & $1.9_{-0.05}^{+0.1}$ &5.75 & 1.737/89\\
\hline 
\end{tabular}
%\end{flushleft}
\begin{flushleft}
$~^a$(z)phabs is the (redshifted)photoelectric absorption model using Balucinska-Church, and McCammon (1992) cross-sections. zpl is redshifted simple power-law model. zbbody is the redshifted blackbody model. z=0.02369 has been used. \\ 
$~^b$Power-law normalization in units of $10^{-5}{\rm~photons~cm^{-2}~s^{-1}~keV^{-1}}$ at 1 keV or blackbody normalization in units of $10^{34}{\rm~erg~s^{-1}~(10~kpc)^{-2}}$. \\ 
$~^c$Observed flux in units of $10^{-13}{\rm ~erg~cm^{-2}~s^{-1}}$ in the energy band of 0.1--2.0~keV. \\
$~^d$Minimum reduced $\chi^{2}$ for $\nu$ degrees of freedom. \\
\end{flushleft}
\end{table*}

Power-law model fits to the spectral data obtained on 1991 June 18, however,
show significant excess absorption over and above the Galactic value 
(see Table 4) when \NH is kept variable.  The best-fit value of 
\NH is about twice of that of the Galactic value thus suggesting an excess 
absorption local to the X-ray source in KUG~1259+280. The best-fit photon index 
is slightly steeper than those obtained previously (see Table 4). In order to 
verify the presence of excess absorption, we fixed the neutral hydrogen 
column density to the Galactic value and fitted the absorbed power-law model. 
Although the photon index becomes flatter, the fit is not acceptable 
($\chi^{2}_{\nu}=2.427$ for 23 degrees of freedom) thus implying the need for an 
excess absorption. To estimate this excess absorption, we introduced an 
additional component for absorption local to KUG~1259+280 in our model, which 
makes the fit acceptable ($\chi^{2}_{\nu}=0.979$ for 22 degrees of freedom) and 
the best-fit value of excess (local to the source) $N_{H}$ 
is $1.0_{-0.3}^{+0.5}\times10^{20}{\rm ~cm^{-2}}$.  
The best-fit photon index in this case is $4.6_{-0.3}^{+0.6}$. 
The best-fit power-law is steeper on June 18 than on June 16 \& 17 (see Table 4).  
The absorbed X-ray flux of KUG~1259+280 in the energy band 0.1--2.0 keV 
is $6.5\times10^{-13}{\rm~erg~cm^{-2}~s^{-1}}$ on 1991 June 18, and the 
corresponding intrinsic X-ray flux is calculated to be 
$7.3\times10^{-12}{\rm~erg~cm^{-2}~s^{-1}}$, based on the best-fit parameters.

In order to further investigate whether the differences in photon indices and \NH  
obtained for the three data sets are indeed real, we have fitted the absorbed 
and redshifted power-law model simultaneously to all the three data sets. 
In this joint fit, the photon indices for the three observations
were tied together and varied to obtain the best-fit. 
Similarly, the \NH and normalizations were also tied together and varied.
The best-fit value of minimum $\chi^{2}_{\nu}$ is 1.119 for 86 degrees of 
freedom which corresponds to a probability of 0.21. The best-fit value 
of the common photon index is $4.25_{-0.22}^{+0.25}$ and 
\NH=$1.4_{-0.2}^{+0.2}\times10^{20}{\rm~cm^{-2}}$  (see Table 4). 
These results indicate that the shape of the spectrum of KUG~1259+280 has 
probably not changed over the three observations. However, \NH column obtained is  
 in excess of the Galactic value suggesting local absorption in all 
three observation. 
To find the excess \NH, we have carried out joint-fit after introducing additional 
absorption local to KUG~1259+280 apart from the fixed Galactic absorption. 
The excess \NH was found to be $5_{-2}^{+2}\times10^{19}{\rm~cm^{-2}}$. 
All other parameters do not change from that obtained in the previous joint-fit. 
The fitted model is shown in Fig. 5, and the confidence contours of 
{\it excess} \NH and photon index are shown in Figure 6. 
A photon index ($\Gamma_{X}$) between 3.8-4.8 and 
an excess absorption between $(0.2-1.0)\times10^{20}{\rm~cm^{-2}}$ are indicated 
for all three observations. The observed X-ray flux of KUG~1259+280 in the energy 
band 0.1--2.0 keV is $6.1\times10^{-13}{\rm~erg~cm^{-2}~s^{-1}}$, and the 
corresponding rest frame intrinsic X-ray luminosity is 
$4.7\times10^{42}{\rm ~erg~s^{-1}}$.

We have also fitted absorbed blackbody model to all three data sets jointly. 
The absorbing column was fixed to the Galactic value for all three data sets. 
The common blackbody temperature was varied to obtain the best-fit. 
Although the fit is not acceptable ($\chi^{2}_\nu= 1.737$ for 89 degrees of freedom),
the temperature ($kT = 0.055{\rm~keV}$) thus inferred reflects the ultra soft nature of KUG~1259+280.

{\begin{figure*}
\vskip 14cm
\includegraphics{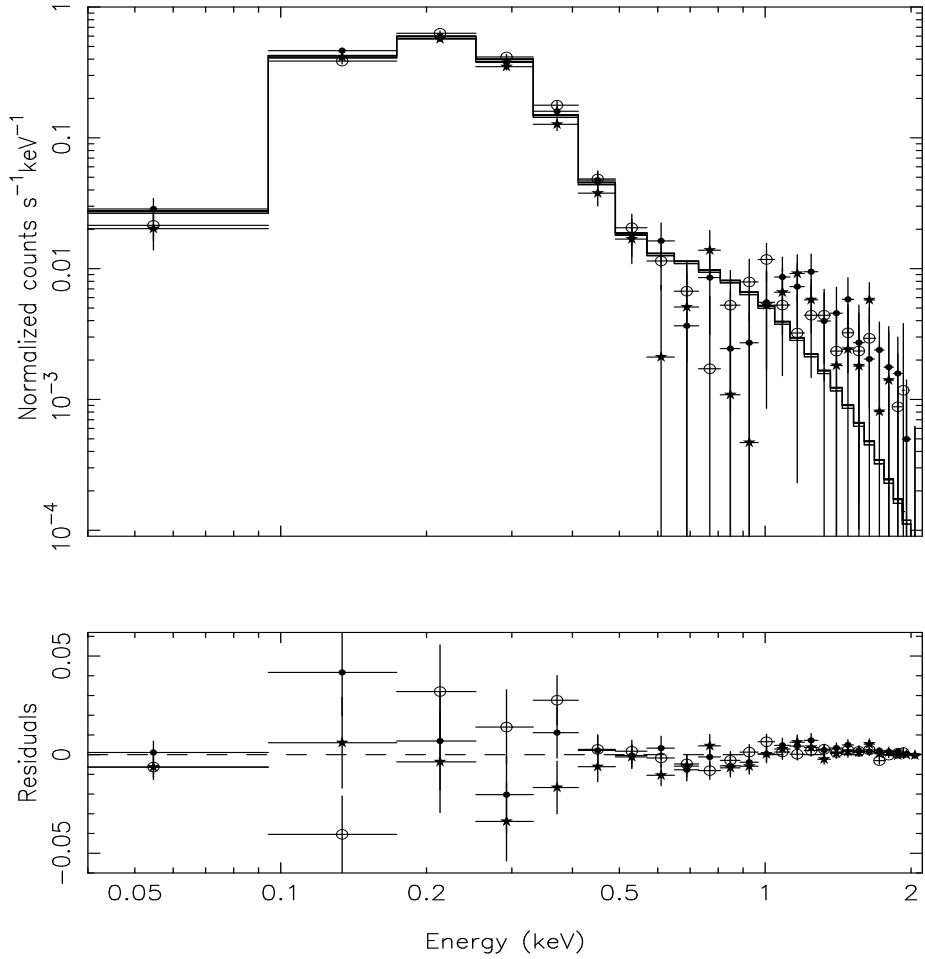}
\caption{
Spectral data and joint model fit to the $ROSAT$ PSPC X-ray spectra of KUG~1259+280. The PSPC spectrum for 1991 June 16 observation is shown as solid circles. The spectrum marked with asterisks is for 1991 June 17 observation. The third spectrum marked with open circles is for 1991 June 18 observation. The jointly fitted power-law model absorbed by Galactic \NH and intrinsic \NH local to KUG~1259+280 is shown with solid line. }
\end{figure*}}

{\begin{figure*}
\vskip 11cm
\includegraphics{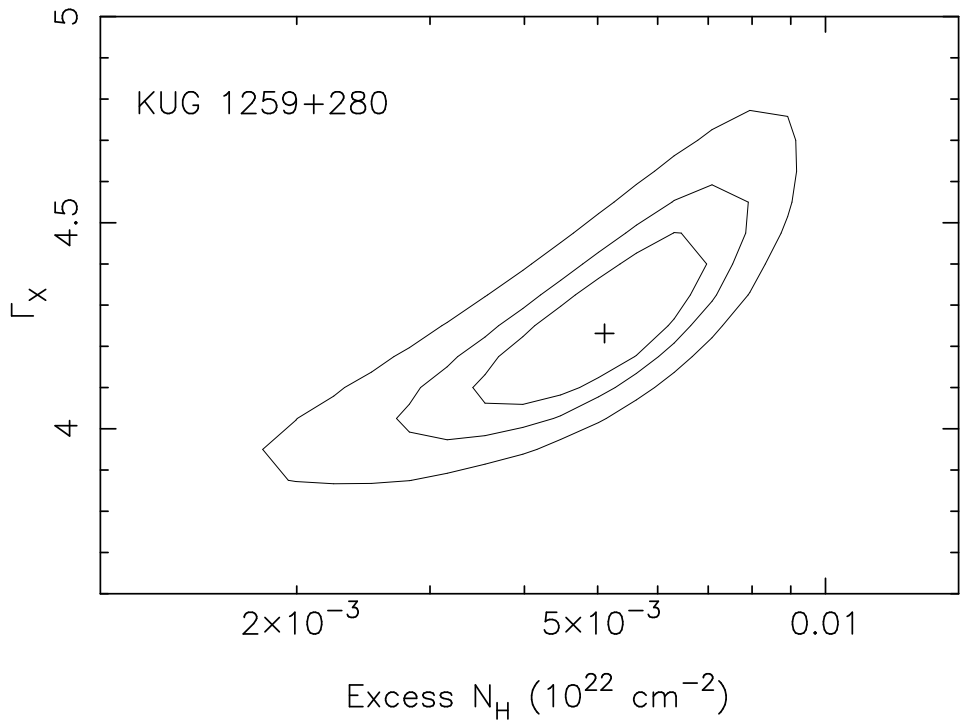}
\caption{Allowed ranges of power-law photon index and `excess' \NH (at the redshift of the source) for $68\%$, $90\%$, and $98\%$ confidence based on counting statistics derived from the joint-fit of the 3 PSPC spectra. The `$+$' marks the best-fit value.}
\end{figure*}}

%We have converted the HRI count rates, listed in Table 2,
%into X-ray flux in the $ROSAT$ HRI
%band of $0.1-2.0$~keV. Since the PSPC spectral model fits to the spectra of 
%KUG~1259+280 have indicated that the shape of the spectrum has not changed over the 
%three observations and there is an excess absorption over the Galactic value, X-ray 
%flux from HRI count rates have been converted using a power-law of  
%photon index ($\Gamma_{X}$) 
%of 4.25 absorbed by total \NH ($1.4\times10^{20}{\rm~cm^{-2}}$) derived 
%by the joint-fit of the three PSPC spectra.
%The observed flux of KUG~1259+280 in the energy band of $0.1-2.0{\rm~keV}$ 
%is found to be $5.1\times10^{-13}{\rm~erg~s^{-1}~cm^{-2}}$ on 1994 June 25 
%and $5.3\times10^{-13}{\rm~erg~s^{-1}~cm^{-2}}$ on 1995 June 5. 
%The corresponding unabsorbed fluxes are $3.4\times10^{-12}{\rm~erg~s^{-1}~cm^{-2}}$ 
%and $3.25\times10^{-12}{\rm~erg~s^{-1}~cm^{-2}}$ on 1994 June 25 and 1995 June 5 
%respectively. The luminosities calculated from unabsorbed fluxes 
%are $3.6\times10^{42}{\rm ~erg~s^{-1}}$ and $3.75\times10^{42}{\rm ~erg~s^{-1}}$ 
%on 1994 June 25 and 1995 June 5, respectively ($z=0.02369$). The observed flux 
%and luminosities of KUG~1259+280 with the HRI are about $25\%$ smaller 
%than those measured from the PSPC. However, it should be noted that the
%measured flux and luminosity values are uncertain by about $10\%$, so the differences 
%in flux or luminosity obtained from the HRI and the PSPC is smaller than $25\%$. 
\subsection {Optical Spectroscopy}
Optical spectra were reduced and analysed using the
IRAF\footnote[7]{IRAF is distributed by the National Optical Astronomy Observatories, which is
operated by  the Association of Universities, Inc. (AURA) under cooperative
agreement with the National Science Foundation. The IRAF version $2.11.3$ was used.}
 software package. Corrections for bias level and response variation of pixels were carried out using the standard procedures for each of the three spectra. 
%Bias level of the CCD was corrected by subtracting the bias  frame from each of the object and the flat frames.
%To correct for the response variation
%along the spatial dimension of the two dimensional spectra, response
%calibration spectrum was created from the flat field frames.
%Several dome flats taken in the same wavelength region and on the same night
%were combined after rejecting the cosmic-ray events. The combined dome
%flat frame was then averaged across the dispersion direction to form a one
%dimensional spectrum, which was smoothed by fitting a low order polynomial
%along the dispersion direction to reduce the noise. The fitted values were then divided
%by the combined dome flat to obtain response calibration spectrum.
%The response correction along the spatial dimension was performed by
%dividing the response calibration spectrum by each of the two dimensional
%object spectrum. 
Three different spectra of KUG~1259+280 were then combined to increase the signal-to-noise ratio. One dimensional spectrum of KUG~1259+280 was then optimally extracted from the two
dimensional spectrum using an aperture of size $4.2\arcsec$ by using the method  of Horne (1986).
The one dimensional object spectrum was wavelength calibrated using
appropriate arc spectrum (He-AR).
 The wavelength calibration was accurate
to $\sim2\AA~$ in all the spectra. Correction for atmospheric extinction was based on the
extinction curve for Kitt Peak National Observatory. However, the correction is not critical as the standard star was observed almost at similar airmass as that of the object. 
Flux calibration was performed
using standard $IRAF$ procedures and using the observed spectrophotometric
standard star HR~4963.  The flux calibrated spectrum of KUG~1259+280 was corrected for Galactic extinction by adopting the extinction law in Cardelli \etal (1989).
Color excess $E_{B-V}$ due to Galactic reddening along the line of sight
of the galaxy KUG~1259+280 was calculated from the neutral hydrogen column density $\NH$
using the relation
\begin{equation}
\NH=5.8\times 10^{21} \times E_{B-V}{\rm ~cm^{-2}}
\end{equation}
(Bohlin, Savage, \& Drake 1978). The dereddened spectrum was then corrected for
Doppler shift (z=0.02369). 
The final reduced spectrum of KUG~1259+280 is shown in Figure 7. The signal-to-noise ratio of the spectrum is $\sim52$ measured at 4500~\AA -- 6600~\AA.
The spectrum of KUG~1259+280  shows
 forbidden emission lines [N II]$\lambda 6548$, [N II]$\lambda 6583$, 
[O III]$\lambda4959$, [O III]$\lambda5007$. 
The Balmer lines H$\beta$, H$\gamma$, and H$\delta$ are 
seen in the absorption. No H$\alpha$ emission is observed in the spectrum. 
The absorption lines of Mg I ($\lambda5176$),
Fe ($\lambda\lambda 5270,5335$), Na I D ($\lambda 5890$), and
Ca II+Ba II blend ($\lambda 6497$) are also present in the spectrum of KUG~1259+280.

{\begin{figure*}
\vskip 13cm
\includegraphics{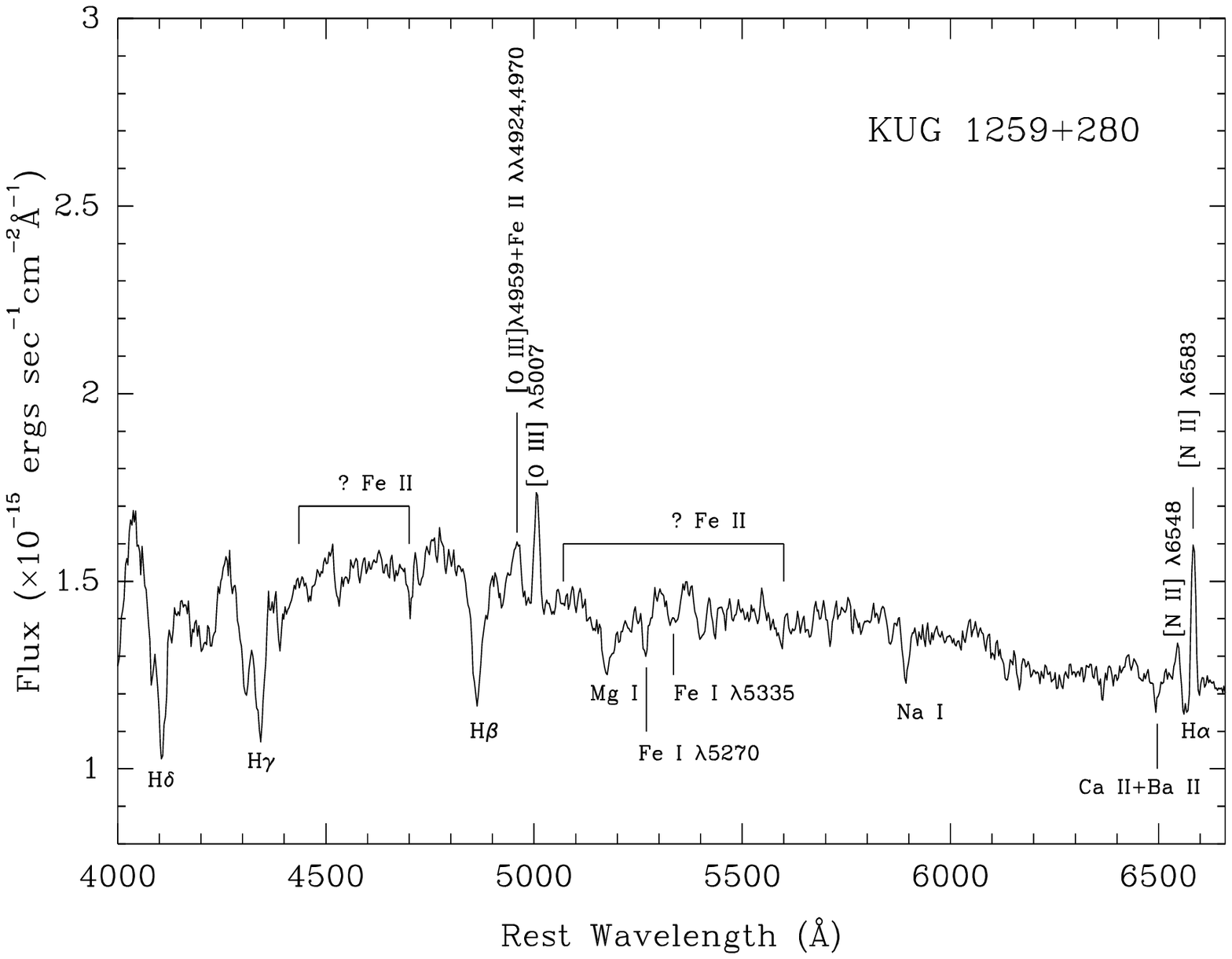}
\caption{Doppler corrected and dereddened optical spectrum of KUG~1259+280. Balmer absorption line H$\beta$, H$\gamma$ and forbidden emission lines e.g., [N II]$\lambda6583$, [O III]$\lambda\lambda4959,5007$ are seen clearly in the spectrum. }

\end{figure*}}

We have fitted Gaussian profiles to the emission 
lines [N II]$\lambda 6583$, [N II]$\lambda 6548$, [O III]$\lambda4959$, 
and [O III]$\lambda5007$ using the profile fitting feature in `splot' 
task within IRAF.  
The spectral line parameters obtained from the fit 
are shown in Table 5. The error quoted for each parameter is the absolute deviation containing
$68.3\%$  of the parameter and were estimated by 100 Monte-Carlo simulations assuming constant Gaussian noise.
The FWHM of [O III]$\lambda4959$ line is about twice of the 
FWHM of the [O III]$\lambda5007$ line and both have similar flux. 
This feature is not an artifact and is confirmed 
by spectrum of the same object taken with the VBT, 
 India but with a poorer signal-to-noise ratio ($\sim30$ at $5000{\rm~\AA}$), albeit slightly higher resolution ($\sim7{\rm~\AA}$). This is not expected and can be attributed 
to the fact that the forbidden line [O III]$\lambda4959$ is blended with the 
Fe II lines at $4924{\rm~\AA}$ and $4970{\rm~\AA}$.
The FWHM of the forbidden lines [N II]$\lambda 6583$, [N II]$\lambda 6548$, 
and [O III]$\lambda5007$ are similar to the instrumental resolution deduced from 
the night sky emission lines. Thus the forbidden lines have not been resolved and 
have FWHMs $\le600{\rm~km~s^{-1}}$. VBT spectrum show FWHM $\simeq300{\rm~km~s^{-1}}$ after removing the effect of instrumental resolution.  
There is some indication for the presence of Fe II blends between 4435--4700${\rm~\AA}$ in 
the blue, and between 5070--5600${\rm~\AA}$. However, the blends seem to be modified 
by the galaxy absorption features e. g. by the presence of Fe $\lambda5335$ absorption 
line usually seen in normal galaxy spectrum (Goudfrooij 1998). To confirm the possible presence of 
Fe II blends higher resolution and higher signal-to-noise spectrum is required. 
Subtraction of a template galaxy spectrum representing the
stellar population of KUG1250+280 from the observed spectrum
will help in confirming the presence of Fe II blends. However,
this requires data with better signal-to-noise ratio and higher resolution
than the data presented here

\begin{table*}
%\centering
\caption{Optical spectral line parameters of KUG~1259+280.}
%\begin{flushleft}
\begin{tabular}{lll}
\hline
Spectral & Line & Equivalent \\
 line & flux & width \\
     & $10^{-15}$~\funit & ${\rm \AA}$  \\
\hline
Emission lines$^1$ \\
${\rm[O~III]}\lambda4959$ &  $3.6\pm 0.46$ & $2.5\pm 0.32$  \\
${\rm[O~III]}\lambda5007$ &  $4.4\pm0.3$ & $3.0\pm0.28$ \\
${\rm[N~II]}\lambda6548$ &  $1.6\pm0.32$ & $1.3\pm0.26$  \\
${\rm[N~II]}\lambda6583$ & $4.7\pm0.24$ & $3.9\pm0.19$  \\
\hline
Absorption lines \\
H$\gamma$     & $8.4\pm0.44$    & $5.8\pm0.31$    \\
H$\beta$      & $9.8\pm0.40$    & $6.5\pm0.26$     \\
Mg~I $\lambda5176$   & $3.1\pm0.39$ & $2.3\pm0.28$   \\
Na~I $\lambda5890$   & $2.2\pm0.33$ &  $1.5\pm0.24$   \\
\hline
\end{tabular}
%\end{flushleft}
\begin{flushleft}
$^1$None of the emission lines have been spectroscopically resolved. [O III]$\lambda4959$ is blended with Fe II $\lambda\lambda5924,4970$. The instrumental resolution (FWHM) is about $600{\rm~km~s^{-1}}$. \\
\end{flushleft}
\end{table*}

\section{Discussion}
Recent optical spectroscopic surveys have been finding a large fraction
($\sim$43\%) of active galactic nuclei among nearby galaxies 
(Ho, Filippenko, \& Sargent 1997a). Nearly 10\% of all nearby galaxies surveyed
by Ho \etal 1997a are found to have Seyfert type nuclei, the rest being LINERS
(low ionization nuclear emission region galaxies) or composite LINER/H II 
nucleus systems.
While the optical emission lines are sensitive to many physical parameters 
e.g., the shape and strength of ionizing continuum arising from nucleus, 
temperature, density, composition and velocities of
line emitting clouds surrounding the nucleus, and dust in the ambient medium, 
X-rays probe the innermost regions of AGNs. Therefore, X-ray
emission properties are more fundamental to infer the nature of galactic nuclei. 
X-ray emission from KUG~1259+280 is found to be unresolved with the $ROSAT$ HRI
observations with a resolution of $\sim4\arcsec$. This suggests that the X-ray
emission from KUG~1259+280 is perhaps mostly of nuclear origin. 
The nuclear nature of X-ray emission is further strengthened by the 
observation of strong and rapid variability detected in the X-ray observations
which also points to a very active nucleus -- an AGN inside KUG~1259+280. 
This is consistent with the point-like appearance of the central region of 
KUG~1259+280 in the optical band in the $HST$ observations with 
$0.1\arcsec$ resolution (CRD).  In the following, we discuss
the observed variability, X-ray and optical spectral
characteristics, and broad-band multiwavelength spectra and 
their implications on models of AGN.

\subsection{X-ray Characteristics \& Variability}
The rest frame intrinsic X-ray luminosity of KUG~1259+280 is in the range of 
$3.6-4.7\times10^{42}{\rm~erg~s^{-1}}$. This is about a factor of 10 higher 
than the X-ray luminosity of low ionization nuclear emission region 
galaxies (LINERs) studied by Komossa, B$\ddot{o}$hringer, \& Huchra (1990) using $ROSAT$. 
All the LINERs in the sample of Komossa \etal (1990) have $L_{X}\le10^{41}{\rm~erg~s^{-1}}$ in the energy band 0.1--2.4 keV. On the other hand, 
Seyfert 1 galaxies studied by Rush \etal (1996) span over 4 orders of 
magnitude in soft X-ray luminosity, from below $10^{42}{\rm~erg~s^{-1}}$ 
to above $10^{46}{\rm~erg~s^{-1}}$ in the same energy band. 
Thus X-ray luminosity of KUG~1259+280 
is similar to that of a low luminosity Seyfert 1 galaxy.

X-ray emission from KUG~1259+280 is highly variable. On 1991 June 18 X-ray 
emission from KUG~1259+280 varied by a factor of $\sim 2.5$ within $\sim$ 
1300~s (Fig. 4). This variability could be due to an outburst or a flare event 
in the nucleus of KUG~1259+280. This kind of time variability is seen in several
AGNs e.g., the low luminosity Seyfert 1 galaxies such as NGC~4051, and 
narrow line Seyfert 1 galaxies such as  IRAS~13224-3809 vary over periods 
of $\sim0.1-1{\rm~hr}$ (Pounds 1979; Lawrence \etal 1985; Boller \etal 1993; 
Singh 1999). On the other hand, rapid variability on the time scales 
less than a day is generally absent in the case of LINERs (Ptak \etal 1998, 
Pellegrini \etal 2000). X-ray emission from KUG~1259+280 is also found to vary by $\sim25\%$ within a year between RASS and PSPC pointed observations.
The large amplitude variation of the nuclear X-ray source in a short time scale 
is possible only if the nucleus is very compact.  Simple light travel time
arguments imply that nuclear X-ray source in KUG~1259+280 is very compact
($size\le1.2\times10^{-5}{\rm~pc}$). Assuming that the soft X-ray excess emission 
originates from a few Schwarzschild radii ($3R_{S}$ or more) away from the 
central massive object, the mass of the central compact object is expected 
to be $\leq 4.4\times10^{7}M_{\sun}$.  A lower limit of the mass of the 
central object can be obtained from the Eddington luminosity.
Since KUG~1259+280 is an ultra-soft X-ray source, it will be reasonable to 
assume that bulk of the luminosity for the nuclear source in KUG~1259+280 is 
in the soft X-ray band. 
Assuming that the Eddington luminosity is about a factor of 5 larger than the 
soft X-ray luminosity in the energy band of $0.1-2.0{\rm~keV}$, the mass of the 
central object is inferred to be at least $2.0\times10^{5}M_{\sun}$. Thus 
the mass of the central compact object is in the range 
of $M\simeq10^{5}-10^{7}M_{\sun}$. The X-ray luminosity of $\sim4\times10^{42}{\rm~erg~s^{-1}}$, small amplitude long term variation, and large amplitude (by a factor of $\sim2.5$) short term variation all indicate that the galaxy harbours an active nucleus.

  The X-ray spectra of KUG~1259+280 obtained from three $ROSAT$ PSPC 
observations are very steep (Table 4). 
A joint-fit has revealed that all three PSPC spectra are well  
represented by a single power-law of $\Gamma_{X}\sim4.25$, with 
an {\it excess} of $\NH=5\times10^{19}$\NHunits over the Galactic value. 
This suggests the presence of an excess absorption above the Galactic value 
during all the three observations. Such an absorber is seen convincingly in 
the observation on 1991 June 18. It is possible that a change in the absorbing column 
density local to KUG~1259+280 could have occurred within $\sim 1{\rm~d}$ 
 as indicated by the individual power-law fits to the PSPC spectra, 
 indicating the presence of moving clouds  close to the central nuclear source. 
The physical state (warm or cold) of the absorbing medium is unclear from 
the present data. However, in the spectrum of 1991 June 18, there is an indication of an absorption 
edge at about 0.8~keV (see residuals in Fig. 5), which is very close to O VII and O VIII edges 
at 0.74 and 0.87 keV commonly seen in $ASCA$ and $ROSAT$ spectra of AGNs (Reynolds 1997). 
But because of the poor signal-to-noise ratio of the data and due to the poor energy 
resolution of the $ROSAT$ PSPC, it is not possible to fit an absorption 
edge.  
Higher signal-to-noise data with better energy resolution will 
be required to confirm the presence of an absorption edge. The short time scale of 
$\sim 1{\rm~d}$ for the change in local \NH also implies that the nuclear 
X-ray source is very compact, similar to those found in AGNs.

\subsection{Optical Characteristics}
Optical spectrum of nuclear region of KUG~1259+280 is not typical of either 
the spectra of normal galaxies or AGNs.  In the optical spectrum of KUG~1259+280,
Balmer emission lines are absent, instead, H$\beta$, H$\gamma$, and H$\delta$ absorption lines are seen. 
H$\alpha$ absorption appears to be almost balanced by emission as was also
pointed out by CRD.  Caldwell \& Rose (1997) have shown that KUG~1259+280 
is a very strong poststarburst galaxy in the Coma cluster, based on the 
presence of enhanced Balmer absorption lines, H$\delta$, H$\gamma$, etc. and 
weaker Ca II K line compared to those in normal galaxies. CRD have pointed out 
that the weak emission line spectrum of KUG~1259+280 is due to nuclear 
activity and not due to residual star formation. From the equivalent 
widths of higher order Balmer absorption lines in the spectrum of KUG~1259, CRD 
have estimated the underlying H$\alpha$ absorption to be $\sim5{\rm~\AA}$ and found 
the ratio of [N II]$\lambda\lambda6548,6583$ to H$\alpha$ emission 
to be slightly greater than 1, which is typical value of LINER emission 
line spectra. It is, however, well known that LINER, Seyfert galaxy, and H II 
region spectra cannot be unambiguously distinguished from one another on the 
basis of any single intensity ratio from any pair of lines. The various 
types of objects with superficially similar emission line spectra can be 
distinguished by using a diagnostic diagram based on pairs 
of emission line intensity ratios (Baldwin, Phillips, \& Terlevich 1981). 
In the case of KUG~1259+280, weak Balmer emission lines arising from the centre 
are dominated by the enhanced absorption lines arising from young stellar 
populations. Assuming Case B recombination (e.g. Osterbrock 1989), 
we estimate the H$\beta$ emission flux to be $\sim2.1\times10^{-15}{\rm ~erg~cm^{-2}~s^{-1}}$ from the equivalent width of H$\alpha$ emission line 
estimated by CRD. Hence the ratio of [O III]$\lambda5007$ to H$\beta$ emission 
is about 2 for KUG~1259+280. Ho, Filippenko, \& Sargent 1997b have defined 
LINER nuclei to be those with $\frac{[O III]\lambda5007}{H\beta}<3.0$, 
and $\frac{[N II]\lambda 6583}{H\alpha} \ge 0.6$.
Based on this definition, the estimated pair of line intensity ratios, 
$\frac{[O III]\lambda5007}{H\beta}\sim2$ and $\frac{[N II]\lambda6583}{H\alpha}\sim 0.8$, 
suggest that nucleus of KUG~1259+280 is, most likely, of LINER type. 
This confirms the conclusion by CRD. However, it should be noted that LINERs show 
the presence of strong [O I]$\lambda6300$ forbidden line in their spectra, but this line is 
not detected in the spectrum of KUG~1259+280. Also, in the spectrum of KUG~1259+280, the forbidden line [O III]$\lambda4959$ is blended with Fe II $\lambda\lambda4924,4970$. The Fe II lines are seen in the spectra of Seyfert galaxies and not in the spectra of LINERs. These raise question regarding the LINER 
nature of KUG~1259+280. 
The lack of Balmer emission lines in the spectrum of KUG~1259+280 can be 
attributed to the poststarburst phenomena.  
To investigate quantitatively the origin of optical emission and absorption lines, higher signal-to-noise and higher resolution optical spectrum as well as UV spectrum is required. 

\subsection{Comparison with Steep Spectrum X-ray AGNs}
It is well known that the steep 
spectrum AGNs generally show optical spectra similar to those of optically identified 
NLS1 galaxies (Boller, Brandt \& Fink 1996). In view of the steep X-ray spectrum of KUG~1259+280, if we assume that the intrinsic 
nuclear optical spectrum is similar to those of NLS1 galaxies modified 
by enhanced stellar absorption, then one would expect that the optical spectrum 
parameters, which are not affected significantly by the stellar absorption, should 
be similar to those of NLS1 galaxies. The interesting parameters are the ratio of 
X-ray to [O III] luminosity, $\frac{f_{x}}{[O III]}$, and slope ($\alpha_{opt}$) of 
the optical spectrum ($f_{\nu}\propto\nu^{-\alpha}$). Grupe \etal (1998) and Grupe \etal (1999) 
have studied soft X-ray and optical properties of 76 steep X-ray spectrum 
AGNs in detail. Based on their measurements, we have calculated $\frac{f_{x}}{[O III]}$ for 71 AGNs  
and found that the ratio, $\frac{f_{x}}{[O III]}$, spans a large 
range (6--3157) with a median value of 399.2. (We have excluded the two objects RXJ0136-35 and a Seyfert 2 galaxy, IC~3599, with $\frac{f_{x}}{[O III]}>15000$ from our analysis. There are only 8 objects with $\frac{f_{x}}{[O III]}>2000$, and only 2 with $\frac{f_{x}}{[O III]}>3200$). 
For KUG~1259+280, $\frac{f_{x}}{[O III]}=156$. 
Thus the ratio, $\frac{f_{x}}{[O III]}$, for KUG~1259+280 is not different from 
those of steep spectrum X-ray AGNs. Since the optical continuum emission of the nucleus 
of KUG~1259+280 is contaminated by the stellar emission as can be seen by the presence of 
strong absorption lines, we have not compared the slope of the observed optical spectrum to those of NLS1 from Grupe \etal (1998). 
The [O III]$\lambda5007$ luminosity of KUG~1259+280 is 
$4.8\times10^{39}{\rm~erg~s^{-1}}$, which is about a factor of about 90 smaller than 
the mean [O III]$\lambda5007$ luminosity of the 3 NLS1s (MS~2340-15, RX~J0439-45, and RX~J2144-39) of Grupe \etal having similar 
soft X-ray slopes as for the KUG~1259. 
This could be due to a dilution of X-ray emission in KUG~1259+280, as the soft X-ray 
luminosity of KUG~1259+280 is smaller by a factor of $\sim750$ than the average soft X-ray luminosity of the 3 AGN in 
the sample of Grupe \etal (1998). The difference in the ratio 
of [O III]$\lambda5007$ to soft X-ray luminosity between KUG~1259+280 
and the 3 AGNs with similar soft X-rays slopes could be either due to the
contribution of [O III] luminosity by H II regions due to hot stars or due to 
differences in the physical parameters of the line emitting clouds, or differences in the shape of the primary radiation in the UV-EUV region. Based on the above comparison, it is likely that KUG~1259+280 is a low 
luminosity version of NLS1 galaxies, where due to the diluted power-law 
continuum, the optical emission lines are weak. Furthermore, the weak permitted lines are 
heavily modified by the enhanced stellar absorption in the poststarburst region. 

\subsection{Spectral Energy Distribution of KUG~1259+280}
In Figure 8, we have plotted far-infrared, optical and soft X-ray flux 
of KUG~1259+280 obtained from various observations. No far-infrared measurements 
have been reported from KUG~1259+280 in the IRAS {\it Point Source Catalogue} (Joint IRAS Working Group 1985). 
The IRAS satellite had, however, scanned the sky containing KUG~1259+280 
during the surveys. We have obtained the IRAS flux after co-adding the available
scan data. The IRAS data were processed using the facilities of IPAC\footnote[8]{The 
IPAC is founded by NASA as part of the IRAS extended mission program under contract to JPL.}.
The measured IRAS flux densities are $0.07\pm0.025{\rm~Jy}$, $0.10\pm0.047{\rm~Jy}$, $0.08\pm0.024{\rm~Jy}$, and $0.41\pm0.105{\rm~Jy}$ at $12$, $25$, $60$, and $100\mu$, respectively.
No radio emission is detected within $15\arcsec$ of KUG~1259+280 in the 
NRAO/VLA and FIRST sky surveys. The spectral energy distribution of KUG~1259+280 is contaminated by stellar emission in the optical region, otherwise it
is similar in shape with those of narrow line Seyfert 1 galaxies reported by Grupe \etal (1999)
and indicates a probable shift in the position of the big blue bump, although not covered in any observation, towards higher energies.

{\begin{figure*}
\vskip 12cm
\includegraphics{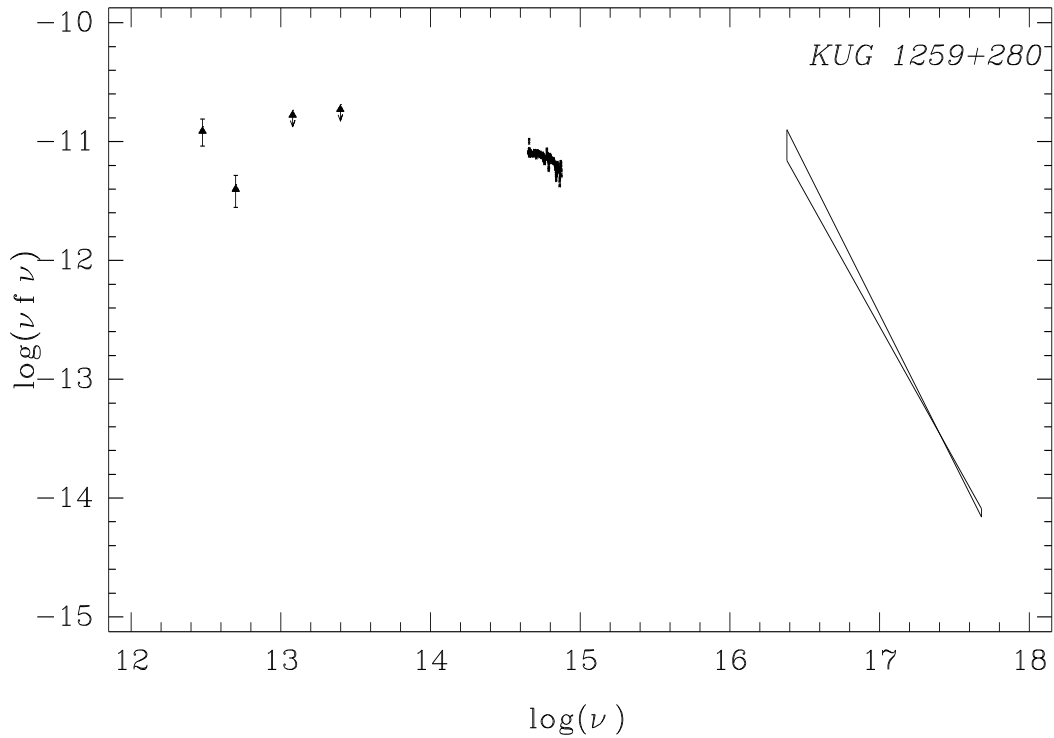}
\caption{Spectral energy distribution of KUG~1259. Arrows indicate $3\sigma$ upper limits.}
\end{figure*}}

\subsection{AGN--poststarburst connection}
As noted earlier, KUG~1259+280 is a nuclear poststarburst galaxy with a 
poststarburst age of $\sim0.5{\rm~Gyr}$ (CRD).  
To build a 
central massive object of mass $10^{6}{\rm~M_{\sun}}$ within KUG~1259+280 after the starburst event, the required average mass accretion rate would be only about $0.002{\rm~M_{\sun}~yr^{-1}}$. 
On the other hand, the Eddington accretion rate for efficiency factor $\eta \simeq 0.1$, and $M=10^{6}{\rm~M_{\sun}}$ is $\dot{M_{E}} \simeq 0.022{\rm~M_{\sun}~yr^{-1}}$. 
Therefore, the required mass accretion rate to build a central massive object of 
mass $\sim10^{6}{\rm M_{\sun}}$ within KUG~1259+280 is about a factor of
10 less than the Eddington accretion rate. Thus it cannot be ruled out that 
the birth of the AGN in KUG~1259+280 took place after or during the starburst event. 
In such a scenario,  
the mass accretion rate onto the young AGN can be expected to be high as the fuel available 
would be enormous. It should be noted that some authors (e.g. Brandt \& Boller 1998) have argued that a high mass accretion rate onto a low-mass ($10^{6}-10^{7}$) black hole can result in AGNs with steep soft X-ray spectrum.

\section {CONCLUSIONS}
(i) X-ray emission from KUG~1259+280 is found to be unresolved by $ROSAT$ HRI 
observations suggesting a point-like nuclear X-ray source in it.

(ii) X-ray luminosity of KUG~1259+280 is estimated to be $\sim3.6-4.7\times10^{42}{\rm~erg~s^{-1}}$ which is similar to that of low luminosity Seyfert nuclei.

(iii) Variability by a factor of $\sim2.5$ within $\sim 1300{\rm~s}$ is detected in
 X-ray emission from KUG~1259+280.
Short time scale of variability strongly suggests an AGN in the galaxy.

(iv) The mass of the central massive object in KUG~1259+280 has been estimated to be in the range $10^{5}-10^{7}M_{\sun}$.

(v) X-ray spectrum of KUG~1259+280 is found to be very steep 
($\Gamma_{X}\sim4.25$) in the $ROSAT$ band 0.1--2.4 keV. 
Individual power-law model fits to 3 PSPC spectra indicate 
a change in intrinsic absorbing column within about a day, while 
a joint-fit of three PSPC spectra indicates excess absorption (total $\NH = 1.4\times10^{20}$\NHunits) over the Galactic value ($\NH = 9.5\times10^{19}$\NHunits).

(vi) The observed optical spectrum of the nuclear region of KUG~1259+280 
is markedly different from that of any class of AGNs. 
The observed spectrum could be understood if the NLR/BLR emission 
line spectrum has been substantially modified by 
enhanced stellar absorption lines. Estimated absorption-free Balmer 
line strengths and observed forbidden line strengths point towards 
LINER nature of KUG~1259+280. However, the nature of X-ray emission points towards NLS1 nature of this galaxy. Presence of Fe II lines supports this picture.

(vii) It is quite possible that the central super-massive object in KUG~1259+280 was formed after or during the starburst event $\sim0.5{\rm~Gyr}$ ago.
%\clearpage
%\newpage
\section {Acknowledgments}
We thank our referee D. Grupe for his comments and suggestions.
This research has made use of $ROSAT$ archival data obtained through the 
High Energy Astrophysics Science Archive Research Center, HEASARC, 
Online Service, provided by the NASA-Goddard Space Flight Center. The 
HEASARC database, the Simbad database, operated at 
CDS, Strasbourg, France;  and the NASA/IPAC Extragalactic Database (NED) which
is operated by the Jet Propulsion Laboratory, Caltech, under contract with
the NASA were used to identify some of the objects.  The PROS software 
package provided by the ROSAT  Science Data Center at Smithsonian 
Astrophysical Observatory, and  the FTOOLS software package provided
by the High Energy Astrophysics Science Archive Research Center
(HEASARC) of NASA's Goddard Space Flight Center have been used.
We thank Booth Hartley for providing IRAS co-added data.

\end{document}